\documentclass[runningheads,a4paper]{llncs}

\usepackage{cmap}
\usepackage[T1]{fontenc}
\usepackage{multirow}
\usepackage{graphicx}
\usepackage[normalem]{ulem}
\usepackage[ngerman,english]{babel}
\addto\extrasenglish{\languageshorthands{ngerman}\useshorthands{"}}

\usepackage[%
rm={oldstyle=false,proportional=true},%
sf={oldstyle=false,proportional=true},%
tt={oldstyle=false,proportional=true,variable=true},%
qt=false%
]{cfr-lm}
\usepackage[math]{blindtext}

\usepackage{cite}

\usepackage{paralist}

\usepackage{csquotes}

\usepackage{microtype}
\usepackage{arabtex}
\usepackage{utf8}
\setcode{utf8}

\usepackage{url}
\makeatletter
\g@addto@macro{\UrlBreaks}{\UrlOrds}
\makeatother
\usepackage{xcolor}

\usepackage{pdfcomment}
\hypersetup{hidelinks,
   colorlinks=true,
   allcolors=black,
   pdfstartview=Fit,
   breaklinks=true}
\usepackage[all]{hypcap}

\usepackage[capitalise,nameinlink]{cleveref}
\crefname{section}{Sect.}{Sect.}
\Crefname{section}{Section}{Sections}

\usepackage{xspace}
\DeclareFontFamily{U}{MnSymbolC}{}
\DeclareSymbolFont{MnSyC}{U}{MnSymbolC}{m}{n}
\DeclareFontShape{U}{MnSymbolC}{m}{n}{
    <-6>  MnSymbolC5
   <6-7>  MnSymbolC6
   <7-8>  MnSymbolC7
   <8-9>  MnSymbolC8
   <9-10> MnSymbolC9
  <10-12> MnSymbolC10
  <12->   MnSymbolC12%
}{}
\DeclareMathSymbol{\powerset}{\mathord}{MnSyC}{180}

\begin{document}

\title{Seminar Users in the Arabic Twitter Sphere}
\titlerunning{Seminar Users in Arabic Twitter Sphere}  % abbreviated title (for running head)
\author{Kareem Darwish\inst{1} \and 
% Sherif Mahmoud\inst{2} \and 
Dimitar Alexandrov\inst{2} \and Preslav Nakov\inst{1} \and Yelena Mejova\inst{1}
}
\authorrunning{Kareem Darwish et al.} % abbreviated author list (for running head)
\tocauthor{Kareem Darwish, Sherif Mahmoud, Dimitar Alexandrov, Preslav Nakov, Yelena Mejova}
\institute{Qatar Computing Research Institute, HBKU, Doha, Qatar\\
\email{\{KDarwish,PNakov,YMejova\}@hbku.edu.qa},\\ 
\and
% Department of Electrical Engineering, Texas A\&M, Doha, Qatar
Sofia University, Bulgaria\\
\email{Dimityr.Alexandrov@gmail.com}
}

\maketitle              % typeset the title of the contribution
\begin{abstract}
We introduce the notion of ``seminar users'', who are social media users engaged in propaganda in support of a political entity.  We develop a framework that can identify such users with 84.4\% precision and 76.1\% recall.  While our dataset is from the Arab region, omitting language-specific features has only a minor impact on classification performance, and thus, our approach could work for detecting seminar users in other parts of the world and in other languages.  We further explored a controversial political topic to observe the prevalence and potential potency of such users.  In our case study, we found that 25\% of the users engaged in the topic are in fact seminar users and their tweets make nearly a third of the on-topic tweets.  Moreover, they are often successful in affecting mainstream discourse with coordinated hashtag campaigns.
\end{abstract}
\section{Introduction}
%{\color{blue} The users we are interested in are influencers, and the persistence of their opinions over time is suspicious, which may suggest intentional opinion manipulation.}
What is the connection between a fishing exhibition in Abu Dhabi and ISIS rhetoric? We encountered this puzzle while analyzing ISIS supporters and opponents on Twitter in 2014 \cite{magdy2016failedrevolutions}, finding, to our surprise, the hashtag % \<\#معرض\_أبوظبي\_للصيد> 
\emph{\#Abu-DhabiFishingExhibition}\footnote{We translate all tweets and hashtags from Arabic to English to ease readability.} to be one of the most discriminating features.  Tracing the accounts that used this hashtag, we found that they had an abnormally high overlap between the tweets they retweeted and the hashtags they used, suggesting they were colluding in some way. However, these accounts did not seem to be political bots \cite{Ferrara2016rise}, which display ``interests'' in multiple subjects, such as the above-mentioned ones.  Neither were they trolls. While their political messages were spirited, they were not  provoking others into conflict \cite{hardaker2010trolling}. They were not selling products or phishing with fake URLs, as spammers would do \cite{benevenuto2010detecting}. However, they were consistent in supporting and promoting the actions of a particular political entity. The closest notion of this behavior in the literature is \emph{astroturfing}, in which users try to give the impression that a grassroots movement is taking place \cite{Ratkiewicz:2011:TMS:1963192.1963301}. Unlike most astroturfing efforts, these users did not mask their identity, and had a persistent political stance over months or years.

\noindent In order to describe users who appear to be real people with persistent political orientation, yet who engage in a coordinated political speech, we use a term originally referring to callers to radio stations who espouse strong political sentiment, namely a ``seminar caller.''\footnote{\url{http://en.wikipedia.org/wiki/Seminar_caller}} The term signifies a specific type of person, who receives instructions in a ``seminar'' on how to deliver talking points on a live talk show effectively, as if sitting in a ``seminar'' instructing them. 
Similarly, Twitter \textit{seminar users} typically attempt to appear like normal individuals, while using specific talking points to promote specific agendas, to manipulate public opinion, and to give the impression of grassroots support.  The operating definition that we use for \emph{seminar users} in this work is the following: ``A politically oriented account acting alone or in a group that is dedicated to the consistent support of a specific entity (government, political party, etc.) or its agenda; the account must not be an official account of the entity.'' A seminar user could be a paid employee of an entity that (s)he actively promotes and could be part of a group of such employees. 
% For example, such an user could be a bot scheduled to retweet tweets from specific accounts with targeted messages at pre-specified or random time intervals, and the bot may be instructed to mix the targeted tweets with retweets from generic accounts that spread generic content such as famous quotes or prayers. For example, a seminar user can be a paid employee of an entity that actively promotes the entity or its messages and could be part of a group of such employees.  The operating definition that we employee in this work for seminar users is the following: ``A politically oriented account acting alone or in a group that is dedicated to the consistent support of a specific entity (government, political party, etc.) or its agenda; the account must not be an official account of the entity.'' 
Incidentally, given our interest in this work is the Arabic Twitter sphere, the term \emph{seminar user} is closely related to the coined Arabic term for such users namely \<لجان إلكترونية>  (\textit{lijan electroniyya}), % ``ljAn Alktrwnyp''\footnote{We use Buckwalter encoding for all Arabic text}, 
which literally translates as ``electronic committees''. 
% Providing a definition for \emph{seminar users} is fairly straight forward.  

Increasingly in the Arab world, seminar users are being employed by governments and opposition parties as a propaganda tool to promote specific agendas, to influence public opinion, or to push policies or future directions. For example, on April 13, 2016, General Sisi of Egypt % warned against using social media as a source, saying 
said he could sway social media agenda using \emph{electronic brigades}.\footnote{\url{http://albedaiah.com/news/2016/04/13/111001} \emph{(in Arabic)}} 
% \textcolor{olive}{I strongly averse to using the term 'president' in referring to Sisi.  He is illegitimate, and I don't recognize him as 'president'.  We need to reword the text without breaking the flow, and without taking a position.}
%Their effect is unmistakable as they carry out deliberate campaigns and are able to initiate hashtags that become trending in a matter of a few hours.  Moreover, they often engage in hashtag hijacking, i.e., they use the hashtags that are being promoted by an opposing party in order to spread their own message instead. 
The activities of such organized users are potentially harmful in a number of ways.  From a data processing perspective, they pollute social media streams, and skew measurements of public opinion.  From a political and social perspective, they may give the impression of grassroots support for questionable or illegal activities such as physically harming opponents. Here are some examples of such activity:
\begin{itemize}{\itemindent=0em}
\item Pro-Sisi %\footnote{Sisi is an Egyptian general who overthrew the democratically-elected government of Egypt in 2013.} % Yelena: This is a politically charged comment, would exclude it
seminar users promoted the hashtag % \<\#متمسكين\_بالسيسي> % ``\#mtmskyn\_bAlsysy'' 
\emph{\#HoldingOnToSisi} in the days leading up to the fifth anniversary of the 2011 Egyptian uprising that overthrew Hosny Mubarak. A sample tweet in this campaign states (translation): ``\#HoldingOnToSisi \#IAmAnElectronicCommitteeForSisi we are proud electronic committees for our country, our president, our military, and our police.''
% RT @reem0ea: #متمسكين_بالسيسي   #انا_لجنه_للسيسي  احنا لجان لبلدنا ولرئيسنا  ولجيشنا وشرطتنا ونفتخر
\item Anti-Sisi seminar users promoted the hashtag %\<\#الشعب\_بيقول\_أحا\_يا\_سيسي> %``\#Al\$Eb\_byqwl\_AHA\_yA\_Sysy'' 
 \emph{\#ThePeopleTellSisiWeStronglyReject(you)} between December 13-16, 2015.  A sample tweet states: ``demonstrate and tell injustice to go away, demonstrate and tell Sisi we strongly reject (you) \#ThePeopleTellSisiWeStronglyReject.''
% \item The two groups may also share a hashtag, competing for its popular definition. For instance, while pro-Sisi seminar users often use Sisi's  %presidential 
% campaign slogan %\<\#تحيا\_مصر> 
% \emph{\#ViveEgypt}, anti-Sisi seminar users use the same hashtag to attack Sisi.
% % RT @9emy1: انزل قول للظلم اتنحى انزل قول للسيسى احا #الشعب_بيقول_احا_يا_سيسي
\end{itemize}

%\textbf{Preslav: 1. I am confused by this example. It does not show hashtag hijacking. It only shows hashtag promoting.}
%\textcolor{red}{Kareem:  the example is NOT for hashtag hijacking}

%\textbf{Preslav: 2. Is it a good idea to use Buckwalter for the Arabic text? Do people in the target conference know what Buckwalter is? Also, it is confusing as it is applied to hashtags and I do not know whether these hashtags are originally written in Arabizi or used Arabic script and now got Buckwaltered.}
%\textcolor{red}{Kareem:  should we put the original Arabic?}

Although the phenomenon of \emph{seminar users} is similar to other phenomena in the field of online political speech, we find no existing definitions that capture this notion in a satisfactory way: these users are not trolls, neither are they astroturfers or bots. In fact, below we will show that state-of-the-art bot detection tools fail to discern these accounts. 

\noindent Thus, the contributions of this work are as follows:

\begin{itemize}
\item We present a method for the automatic detection of seminar users that engage in political discourse in the Arab world.  We manually label users who are engaged in political discourse related to Egypt, United Arab Emirates, Saudi Arabia, and Yemen. Using these identified seminar users, we build a classification model using a variety of content-based and network features. 
\item As a case study, we use the classifier to identify and study pro- and anti-Sisi seminar users.  Specifically, we examine (\emph{i})~how much influence such users have on normal users, (\emph{ii})~whether seminar users are colluding, and (\emph{iii})~whether there is any interaction between seminar users from opposing sides.
\end{itemize}
% \vspace{-20pt}

The remainder of this paper is organized as follows. First, we discuss some related work. Then, we present our experiments in finding seminar users. %: from data creation to methodology and evaluation. 
This is followed by a case study of pro- and anti-Sisi seminar users.
Finally, we conclude with a general discussion and possible directions for future work.

\section{Related Work}
\label{sec:related}

The behavior of \emph{seminar users} is closely related to other manipulative and potentially disruptive practices in social media, including discussion trolling, political astroturfing, sockpuppeting, and use of Internet water army, which we outline below. We further discuss some recent work that analyzes political speech in Arabic.

The promise of social media to democratize content creation \cite{kaplan2010users} %has been %accompanied by the malicious use of this new medium. 
has also been accompanied by many malicious attempts to spread misleading information over this new medium. News community forums in particular saw the rise and proliferation of fake news \cite{Hardalov2016}, aggressiveness \cite{moore2012anonymity}, and trolling \cite{cole2015s}. The latter often is understood to concern malicious online behavior that is intended to disrupt interactions, to aggravate interacting partners, and to lure them into fruitless argumentation in order to disrupt online interactions and communication \cite{Chen:2013:BIW:2492517.2492637}. 
Thus, Twitter has taken measures to suspend users who are recognized to be malicious\cite{Wei:2015:FTS:2808797.2809316}. Nevertheless, more sophisticated malicious profiles such as opinion manipulation \emph{trolls} (paid \cite{Mihaylov2015ExposingPO} or just perceived \cite{Mihaylov2015FindingOM}), \emph{sockpuppets} \cite{Maity:2017:DSS:3022198.3026360,Bu:2013:SPD:2400768.2401510,Kumar:2017:AMS:3038912.3052677,Liu:2016:SGD:2872563.2872566}, and \emph{Internet water army} \cite{Chen:2013:BIW:2492517.2492637} are still barely detectable.
% \vspace{-10pt}

\paragraph{Sockpuppets} are people who assume a false identity in an Internet community and then speak to/about themselves while pretending to be another person. The term has also been used to refer to opinion manipulation, e.g., in Wikipedia \cite{SolorioHM14}.
Sockpuppets have been identified using authorship-identification techniques and link analysis \cite{Bu:2013:SPD:2400768.2401510}.
It has been also shown that sockpuppets differ from ordinary users in their posting behavior, linguistic traits, and social network structure; moreover, sockpuppets tend to start fewer discussions, write shorter posts, use more personal pronouns such as ``I'', and have more clustered ego-networks \cite{Kumar:2017:AMS:3038912.3052677}.
Furthermore, gangs of sockpuppets controlled by a single person can be identified by the similarity of sentiment orientation toward topics based on their posted comments \cite{Liu:2016:SGD:2872563.2872566}.
Unlike \emph{sockpuppets}, \emph{seminar users} do not try to hide their identity.  

\paragraph{Internet water army} is a literal translation of the Chinese term \emph{wangluo shuijun}, which is a metaphor for a large number of people who are well organized to flood the Internet with purposeful comments and articles. Internet water army is allegedly used in China by the government (known also as \emph{50 Cent Party} as people involved in the campaign were allegedly paid about 50 cents per post) as well as by private organizations. 
Chen \& al. \cite{Chen:2013:BIW:2492517.2492637} used semantic analysis and some non-semantic features such as percentage of replies among the posts, average time between posts, number of active days, and number of news posts a user has commented on.
Unlike \emph{Internet water army} users, \emph{seminar users} could act alone and are consistent in their support for a given entity over a long period of time.

\paragraph{Trolling behavior} is present in all kinds of online media: online magazines \cite{binns2012don}, social networking sites \cite{cole2015s}, online computer games \cite{thacker2012exploratory}, online encyclopedia \cite{shachaf2010beyond}, government e-petition pages \cite{virkar2016trolls}, online newspapers \cite{ruiz2011public}, etc. Trolling can be dangerous, as it can increase the risk of suicidal behavior and self-harm amongst the users it is targeted at \cite{slee2016school}.
%Thus, the development of algorithms and methods for reliable and automatic detection of trolls and trolling events have become an important research direction recently.}
\emph{Troll detection} has been addressed using semantic analysis \cite{cambria2010not}, domain-adapted sentiment analysis \cite{Seah2015}, various lexico-syntactic features about user writing style and structure 
%and specific cyber-bullying content
%as features in order to predict the probability of sending out offensive content
\cite{chen2012detecting}, as well as graph-based approaches over signed social networks \cite{Ortega20122884}. 
There have been also studies on general troll behavior \cite{buckels2014trolls} and cyber-bullying \cite{sarna2015content}, 
as well as on linking fake troll profiles to real users \cite{galan2015supervised}.
The term \emph{troll} is often used in popular culture to designate users who engage in opinion manipulation in social media and Web forums; it has been also shown that users who have been called a \emph{troll} by several different people have common characteristics with paid opinion manipulators \cite{mihaylov-nakov:2016:P16-2}.
%Some studies related to cyber-bullying have already been applied in real settings in order to detect and stop cyberbullying in elementary school by using a supervised machine learning algorithm that links the fake profiles to real ones on the same social media \cite{galan2015supervised}. 
Although similar in their ability to influence discourse, seminar users do not employ as disruptive language as trolls. Instead, they push a consistent political message, which is more closely aligned to \emph{astroturfing}, and more precisely a kind of an \emph{Internet water army}\cite{Chen:2013:BIW:2492517.2492637}. 
% \vspace{-10pt}

\paragraph{Astroturfing.} Named after a brand of fake plastic grass, astroturfing is an effort to simulate a fake political grassroots movement. Recent preoccupation with it has been motivated by strong interest from political science, and research methods driven by the presence of massive streams of microblogging data, largely centering around the Twitter accounts involved. Since Ratkiewicz et al. proposed a system to detect astroturfing called Truthy \cite{Ratkiewicz:2011:TMS:1963192.1963301}, studies have shown that people can be poor judges of a tweet's credibility \cite{morris2012tweeting}. Subsequently, tools to automatically assess the credibility of tweets and their origins spanned both political news \cite{castillo2013predicting} and disaster response \cite{gupta2013faking}, as well as for the related tasks of stance classification \cite{zubiaga2016stance} and contradiction detection in rumors \cite{lendvai2016contradiction}. Finally, Lukasik et al. \cite{lukasik-cohn-bontcheva:2015:ACL-IJCNLP} and Ma et al. \cite{Ma:2015:DRU} used temporal patterns to detect rumors,
%and to predict their frequency,
and Zubiaga et al. \cite{PlosONE:2016} focused on conversational threads.
% \vspace{-10pt}
\paragraph{Identification of malicious accounts} in social networks reaches beyond fake political grassroots movements, and includes detecting spam accounts \cite{almaatouq2016if,mccord2011spam}, fake accounts \cite{cresci2015fame}, compromised accounts and phishing accounts \cite{adewole2017malicious}. Fake profile detection has also been studied in the context of cyber-bullying \cite{galan2014supervised}. 

\paragraph{Web spam detection} is another related problem, which has been addressed as a text classification problem \cite{sebastiani2002machine}, e.g., using spam keyword spotting \cite{dave2003mining}, lexical affinity of arbitrary words to spam content \cite{hu2004mining}, frequency of punctuation and word co-occurrence \cite{li2006combining}. Yet another related problem is that of detecting racist/radicalized posts, e.g., on the Tumblr micro-blogging website \cite{agarwal2017characterizing}. 

% \vspace{-10pt}
\paragraph{Predicting online extremism} is a topic, which involves fighting with a special type of malicious users, very similar to the seminar users group. The goal in \cite{ferrara2016predicting} is to find Twitter accounts used by ISIS members and to report them to Twitter. After the manual annotation of some ISIS supporters, the authors used a classifier (based on logistic regression and random forests) to predict whether a regular user would retweet a message posted by an ISIS account \cite{ferrara2016predicting}. 
% The third task presents likely the most difficult challenge:
% predicting whether a regular user will engage into interactions
% with extremists. A positive instance of interaction is
% represented by a regular user replying to a contact initiated
% by an ISIS account.\cite{ferrara2016predicting}

\paragraph{Social bot detection} involves detecting accounts that are programmatically controlled to produce content and to interact with other users. Social bots may spread messages, ads, or propaganda while trying to mimic human behavior \cite{varol2017online}.  Varol et al. \cite{varol2017online} detected bots using profile, friends, network, temporal, content, and sentiment features. For detecting seminar users, we use features such as lexical diversity and sentiment, which overlap with their features. For social bots in the Arab world, Morstatter et al. \cite{Morstatter:2016:botDetection} looked for keywords that relate to the Arab Spring in Libya. They used several features such as retweet and URL sharing frequencies, tweet length, time between tweets, and topic distribution. Abokhodair et al. \cite{abokhodair2015dissecting} looked at the behavior of a social botnet related to the ongoing civil war in Syria and the propensity of the botnet in influencing the mainstream discourse. 
%The main concept in this paper is to focus not only on the precision, but also on the F1 and recall metrics. 
% The bot detection approaches include Honeypot network, heuristics and topic modeling, based on LDA. 
%Two datasets have been used: one labeled by Twitter, which specifies whether the users from the set are active, suspended or deleted, and another one based on a honeypot approach\cite{Morstatter:2016:botDetection}

Unlike the work above, here we focus on a particular kind of malicious users, i.e., \emph{seminar users}, which show dedicated support to a particular political entity, acting alone or as part of an organized group.

\section{Identifying Seminar Users}
\label{sec:findingseminarusers}

% { \color{blue} Seminar users skew our view of Twitter conversations by overpowering the more ``real'' users (show an example, refer to vocal minority, silent majority \cite{mustafaraj2011vocal}) -- also look at early work of Moscovici:  https://www.simplypsychology.org/minority-influence.html}
\subsection{Data}
\label{id-data-section}
We collected Arabic tweets between December 1, 2015 and January 25, 2016.  This period is interesting as it constitutes the days leading up to the fifth anniversary of the January 25, 2011 uprising in Egypt, which is part of the Arab Spring and led to the ouster of Hosny Mubarak from power. %\textbf{Preslav: Uprising of who against whom? This is probably not so clear outside Egypt.}. 
Though the dates are unlikely to affect the first part of the work concerning the automated detection of seminar users, it would likely make subsequent analysis of specific pro- and anti-Sisi seminar users more interesting. We collected the tweets using the Twitter4j Java interface for the Twitter streaming API, where we searched using the query ``lang:ar'' to get Arabic tweets.  %We used multiple proxies to overcome the rate limitations of Twitter. 
Eventually, we accumulated 417 million tweets with an average of 7.4 million tweets per day over 56 days. 
% \textbf{Preslav: Maybe no need to say about the multiple proxies as this probably violates Twitter's terms and conditions. Also, maybe no need to say that Twitter4j was used: does it matter?}
We manually labeled 150 users as seminar users vs. non-seminar users.  %We did not tag random users.  Instead 
To label them, we picked users who discussed potentially politically polarizing issues about Egypt, United Arab Emirates (UAE), Yemen, and the Kingdom of Saudi Arabia (KSA). 

\noindent Specifically, we randomly picked 2,000 users (500 for each country) who mentioned General Sisi of Egypt, Mohamed bin Zayed (vice president of UAE), Yemen, and Saudi Arabia in their tweets. Next, an annotator who is privy to the politics of the Middle East manually examined the tweets of the 2,000 users in random order until he tagged 150 users as fitting our definition of seminar users or not.  In the interest of data cleanliness, the annotator labeled users who were clearly seminar users or were clearly not, and he excluded borderline (ambiguous) users where he could not make a clear-cut decision.  The annotator was asked to keep the dataset balanced with a comparable number of seminar and non-seminar users.  Of the 150 users, 71 were seminar users and 79 were not. 
% \textbf{Preslav: This is confusing. It reads as if there are 150 seminars users out of 2000 and 71 of them are trolls and 79 are not. As if seminar users can be either troll or not. This mixes terminology and definitions. It is also not clear what happened to the the remaining 2000-150=1850 users. How were the 150 selected? Let us not talk about trolls here, just about seminar users.}
To ensure the quality of the annotation, another annotator was asked to independently tag a subset of the tagged users containing 35 users and both annotators agreed on 32 users (91.4\% agreement, with a Cohen's Kappa of 0.83).
% \textbf{Preslav: We need to calculate the Kappa statistic between the two annotators.}

\begin{figure}[ht]
\centering
\includegraphics [width=.70\linewidth]{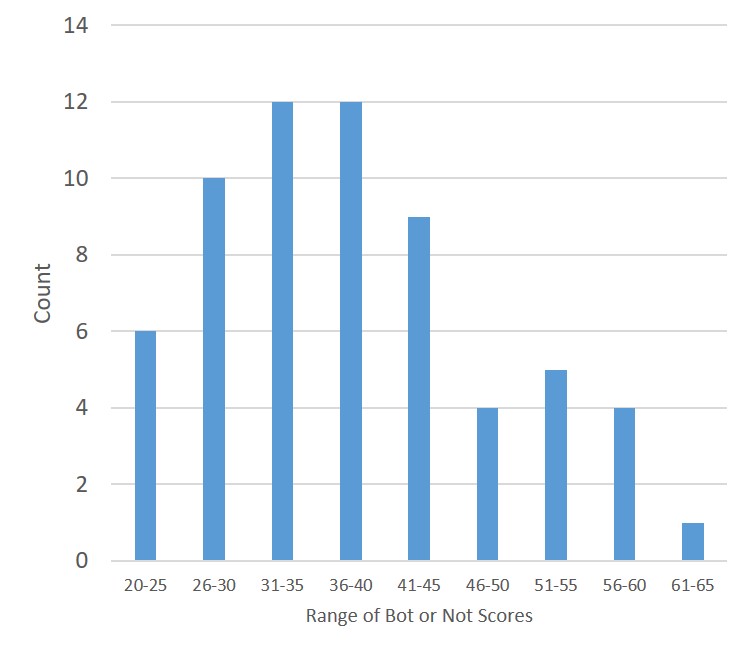}
\caption{Distribution of BotOrNot scores for seminar users in our dataset}
\label{botornot}
\end{figure}

Tagging users as seminar users is quite different from tagging them as bots.  To demonstrate the difference, we automatically tagged all seminar users from our dataset using BotOrNot,\footnote{\url{http://truthy.indiana.edu/botornot/}} which gives a score between 0 and 100 to each user, with 100 signifying the most extreme bot-like behavior \cite{davis2016botornot}.  Figure \ref{botornot} shows the scores for the seminar users (after excluding 16 accounts that were deleted, protected, or suspended).  As the figure shows, BotOrNot reckoned that most of the accounts were not bot-like, with only one user getting the highest score of 61 for any user in our dataset.  This may indicate that the accounts are being managed by humans. However, this does not preclude the possibility that some user might manage more than one account. 

% \textbf{Should we have a table that summarizes the data? It should show how many seminar vs. non-seminar users we have, and also how many of the seminar users are pro- vs anti-Sisi. Also, it should show how many were automatically generated using label propagation.}

\subsection{Detecting Seminar Users}
\subsubsection{Methodology}
Given our set of 150 users that were tagged as seminar users or not, we trained an SVM classifier with a radial basis function kernel\footnote{We used the SVM$^{Light}$ implementation available from \url{http://svmlight.joachims.org/}} using a variety of features:\\
% \begin{itemize}
$\bullet$ (Interaction features) The relative use of tweet interaction elements.  Specifically, the percentage of tweets containing retweets, URLs, hashtags, mentions, and embedded images. The rationale for these features is that seminar users may engage heavily in the promotion of hashtags or memes. (5 features) \\
$\bullet$ (Diversity features) The diversity of content, where seminar users may concentrate on specific topics and use a smaller vocabulary.  We measured
  \begin{itemize} %[leftmargin=5pt]
  \item the percentage of tweets containing the top $n$ retweeted accounts or mentioned accounts, where $n$ was set to 1, 3, 10, and 20 (8 features);
  \item the percentage of tweets containing the top $n$ used hashtags for each user, where $n$ was set to 5 and 15 (2 features);
  \item the percentage of tweets containing the top $n$ used words (excluding stopwords) for each user, where $n$ was set to 10, 30, and 50. We performed basic Arabic letter normalization \cite{darwish2012language} and English case folding. (3 features)
  % \textbf{Preslav: I imagine that stopwords were excluded? This has to be said.}
  \end{itemize}
$\bullet$ (Style features) The style of language, where seminar users are more likely to praise the party they support and malign opponents.  We used the following:
	\begin{itemize}
	\item The percentage of tweets containing sentiment words. We used the sentiment lexicon of 3,982 words described in \cite{muhammad2012samar}. (1 feature)
    \item The percentage of tweets containing vulgar or combative words.  We used a word list containing Arabic vulgar and obscene words %that was developed by Mubarak et al. \cite{mubarak2017vulgar}, %constructed a list of vulgar and combative words in the following way. We created a set of obscene words to work as our seed list. To create the list, we searched Twitter for some patterns that are usually used in offensive communications, such as: ``You'' and ``son of''. The words appearing after these patterns were
% then collected and manually assessed for being vulgar/combative or not. In addition, we added hashtags that are used to filter out pornographic tweets in an online Twitter aggregator called \emph{TweetMogaz} (Magdy, 2013; Elsawy et al., 2014). \textbf{Preslav: we need actual cite\{\} here.} Overall, the list
containing 3,840 words and phrases such as ``son of a dog'' and ``filthy'' \cite{mubarak2017vulgar}. (1 feature)
	\end{itemize}
% \end{itemize}
We avoided using user similarity features, as such features are computationally expensive.  The advantage of our proposed features is that each user can be classified in isolation, without considering the other users, which makes classification computationally efficient.  Also, since many users subscribe to services that post mostly religious tweets on their behalf, we assembled a list of the most popular such services and filtered out all tweets that they generate. %, where the tweets they generate include their URL.  
The services we excluded are \url{du3a.org}, \url{ghared.com}, \url{7asnat.com}, \url{mezani.net}, \url{d3waapp.org}, \url{zad-muslim.com}, and \url{rtw8.com}.  We also excluded tweets containing the hashtags \#quran and \#hadith, or that were retweeted from specific accounts 
%such as \@\url{alsa7e7a}, \@\url{tasbihistighfar}, \@\url{W888F}, \@\url{alathkaar}, and \@\url{alathkaar\_E}, 
that specifically indicate religious content.

\subsubsection{Evaluation}
Due to the limited number of examples in our dataset,  we used leave-one-out cross-validation for training and testing, where we held each of the users in our set out for testing and trained on the remaining users.  
% \textbf{Preslav: I have not seen ``hold-one-out'' before; I have seen ``leave-one-out cross-validation''.}
%This is essentially the same as cross-validation where the number of folds equals the number of items in our dataset.  
We experimented with using the Interaction, Diversity, and Style features separately and in combination. 

\noindent Table \ref{seminarResultsRBF} reports the classification results using precision (P), recall (R), and F1-measure (F1).  We can see that the interaction features are the most important ones.  While the combination of all feature types yields the best results, using a combination of interaction and diversity features only, which are language-independent, slightly degrades the performance. This gives us hope that the classifier may work effectively for users in other regions without the need for language-specific resources. Overall, the results show that we can identify seminar users with relatively high precision and recall.  %\textcolor{red}{Yelena: A few words on how well this is doing compared to, say, bot detection? or other related tasks?}
\begin{table}
\begin{center}
\begin{tabular}{r|c|c|c|c|c|c|c}
           & \multicolumn{3}{c}{Seminar} & \multicolumn{3}{|c|}{Normal} & Macro \\ %\hline
Features & P & R & F1 & P & R & F1 & F1\\ \hline
Interaction & 77.2 & 62.0 & 68.8 & 71.0 & 83.5 & 76.7 & 72.8\\
Diversity & 73.2 & 57.7 & 64.6 & 68.1 & 81.0 & 74.0 & 69.3\\
Style & 0.0 & 0.0 & 0.0 & 52.7 & \textbf{100.0} & 69.0 & 34.5\\
Interaction + Diversity & 83.9 & 73.2 & 78.2 & 	78.4 & 87.3 & 82.6 & 80.4 \\ \hline
All & \textbf{84.4} & \textbf{76.1} & \textbf{80.0} & \textbf{80.2} & 87.2 & \textbf{83.6} & \textbf{80.9} \\
\end{tabular}
\label{seminarResultsRBF}
\caption{Classification results for seminar users and normal (non-seminar) users including precision (P), recall (R), F1-measure (F1), and Macro F1-measure.
%\color{red}{\textbf{Preslav: can we put also a baseline?}}
}
\end{center}
\end{table}
% Even using Style features leads to better overall results, despite the fact that using Style features alone leads to a recall of zero. \textbf{Preslav: the last sentence is confusing.}
% \begin{table}
% \begin{center}
% \begin{tabular}{r|c|c|c|c|c|c}
% 		 & \multicolumn{3}{c}{Seminar} & \multicolumn{3}{|c}{Normal} \\ \hline
% Features & P & R & F & P & R & F \\ \hline
% Interaction & 68.8 & 77.4 & 72.8 & 77.1 & 68.4 & 72.5 \\
% Diversity & 68.5 & 52.1 & 59.2 & 64.6 & 78.5 & 70.1\\
% Style & 0.0 & 0.0 & 0.0 & 52.7 & 100.0 & 69.0 \\
% Interaction + Diversity & 82.6 & 53.5 & 65.0 & 	68.3 & 89.9 & 77.6 \\ \hline
% All & 85.1 & 56.3 & 67.8 & 69.9 & 91.1 & 79.1 \\
% \end{tabular}
% \label{seminarResults}
% \caption{Classification results for seminar users and normal (non-seminar) users including precision (P), recall (R), and F1-measure (F)}
% \end{center}
% \end{table}
% \vspace{-10pt}

\section{Pro- and Anti-Sisi Seminar Users: Case Study}
\label{sec:casestudy}

\subsection{Data}
For the second part of the study, we focus on the behavior of pro- and anti-Sisi seminar users.  On July 3, 2013, General Sisi used the military to overthrow the democratically elected government that transpired after January 25, 2011 Egyptian uprising. He was named president a year later. Sisi is a divisive figure in Egyptian politics.  We identified 9,506 users who mentioned Sisi ten or more times in our aforementioned dataset  containing tweets that are collected between December 1, 2015 and January 25, 2016.  To label users as pro- or anti-Sisi, we used label propagation.  We manually tagged an initial set of 100 users. 
% \textbf{Preslav: Why not tag as pro/anti-Sisi all 150 initial users? Hmm, but only 71 of them are seminar users. How come now you there are 100 known seminar users? If we have them now why not in the first dataset? Also, what is the initial proportion of pro- vs. anti-Sisi seminar users? And what is the final proportion? Maybe show this proportion for each of the three iterations in a table or a graph?} 
Then, we automatically tagged all the tweets mentioning Sisi for those 100 users using their user tags.  For example, all tweets mentioning Sisi and posted by pro-Sisi users would be tagged as pro-Sisi.  It is not unreasonable to do so as people's opinions generally remain stable over extended periods of time.  In particular, we assumed that the overwhelming majority of users would not change their opinion over the span of 56 days.  We further assumed that unlabeled users who retweeted consistently pro-Sisi tweets should be pro-Sisi, and likewise those who consistently retweeted anti-Sisi tweets should be anti-Sisi.  Thus, we were able to tag 2,743 more users.  We repeated this process for three iterations, which ultimately yielded 7,427 tagged users. To check the reliability of label propagation on our data, we took a random sample of 100 users and we labeled them for stance manually and also automatically, obtaining the same label for 99 users.  
% \textcolor{red}{\textbf{Preslav: Do we need to manually check a random sample of these 7,532 users to see how well label propagation actually worked? Otherwise, we would be attacked that we make too many assumptions and we never check.}}
We applied our seminar user detector on the 7,427 users that we identified as pro- and anti-Sisi.  As a result, we identified 1,839 users as seminar users, including 492 who were pro-Sisi and 1,347 who were anti-Sisi.
In order to verify the efficacy of the classifier on the new data, we randomly sampled 50 users from the set of 7,427 users, and we asked our annotator to manually and independently label them as seminar users or not. The annotator labeled 20 users as seminar users, 29 as normal users, and 1 as a spammer. Our classifier labeled 14 (out of the 50) users as seminar users, of which 11 were in fact seminar users (Precision = 0.78; Recall = 0.55). Unlike the training data, not all users were clearly discernible as seminar users or not. The annotator labeled all ambiguous users (8 users) using his best guess. This likely contributed to lower results over the sample compared to the aforementioned classification results.

Table \ref{proAntiSisiStats} summarizes the number of users for each group and other basic statistics about them.  The summary shows that the ratio of seminar users to normal users is 1:1.7 and 1:3.5 for pro- and anti-Sisi users, respectively. It is noteworthy that pro-Sisi seminar users authored more tweets than pro-Sisi non-seminar users. Conversely, normal anti-Sisi users produced more than twice as many tweets as anti-Sisi seminar users.  Seminar users produced on average 60\% more tweets per user, and they tweeted more on average about Sisi. 

\begin{table}[ht]
\begin{center}
\begin{tabular}{l|r|r|r|r}
			& \multicolumn{2}{c}{Seminar Users} & \multicolumn{2}{|c}{Normal Users} \\ \hline
Stance     	& Pro-Sisi	& Anti-Sisi & Pro-Sisi	& Anti-Sisi \\ \hline
Users & 492	& 1,347 & 846	& 4,748 \\
Tweets & 860,085	& 4,200,619 & 847,383	& 9,057,758 \\
Avg tweets/user & 1,748 & 3,118 & 1,001 & 1,908 \\
% Tweets mentioning Sisi & 22,932 (2.7\%) & 65,478 (1.6\%) & 22,932 (2.7\%) & 65,478 (1.6\%) \\
Tweets mentioning Sisi & 22,932 & 65,478 & 34,026 & 146,762 \\
Avg tweets mentioning Sisi/user & 46.6 & 48.6 & 40.2 & 30.9 \\
\end{tabular}
\caption{Stats of pro- and anti-Sisi users and tweets.}
\label{proAntiSisiStats}
\end{center}
\end{table}
% \vspace{-30pt}

\subsection{Characteristics of Pro- and Anti-Sisi Seminar Users}

\begin{figure}
\centering
\includegraphics[width=\linewidth]{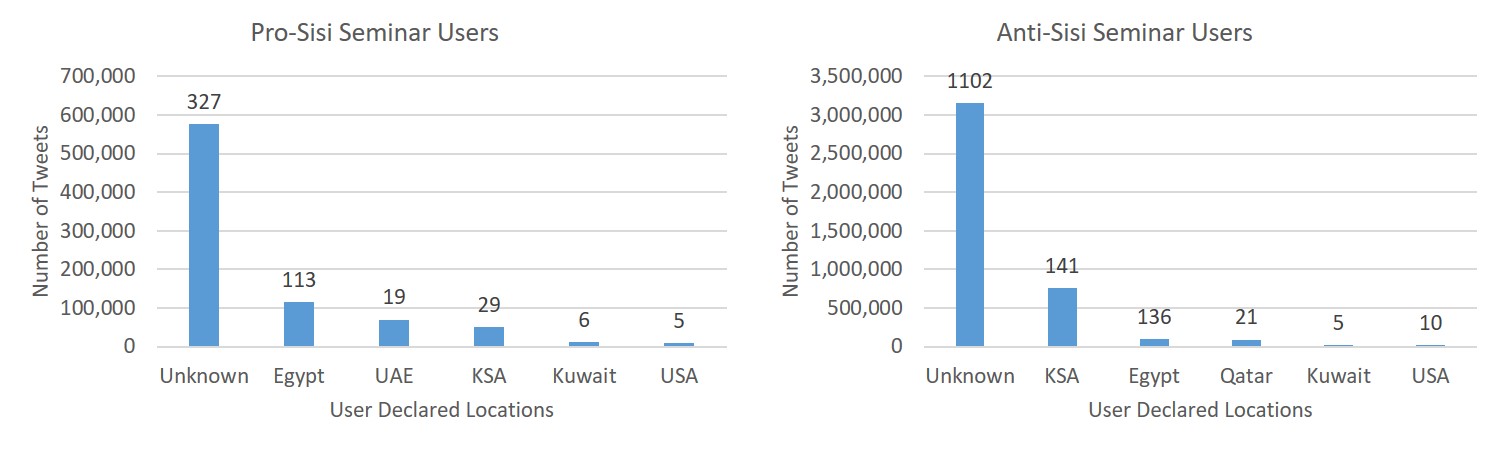}
\caption{Top-5 self-declared user locations of pro- and anti-Sisi seminar users. Values at the top of the bars indicate the number of users.}
\label{sisiGeoLoc}
\end{figure}

Figure \ref{sisiGeoLoc} shows the top-5 self-declared user locations producing the highest number of pro- and anti-Sisi tweets.  We can see that most users did not declare their location.  Moreover, even though there were far fewer pro-Sisi users from UAE compared to Egypt, they produced five times as many tweets. % We manually inspected the accounts and 6 of the users were clearly from UAE and 1 was clearly Egyptian.
Similarly for anti-Sisi users, there were nearly six times as many users from Egypt than from KSA, but the number of tweets from Egypt were only 19\% more than those from KSA.  This may indicate that Egyptian seminar users are less active than non-Egyptian ones.  Table \ref{topRetweetedAccounts} lists the top retweeted accounts by both groups.  As the table shows, for both groups, the majority of these accounts were not from Egypt, and UAE and KSA users account for most pro-Sisi seminar users.

% \begin{table}[ht]
% \begin{center}
% \scriptsize
% \begin{tabular}{l|r||l|r}
% \multicolumn{2}{c|}{Pro-Sisi} & \multicolumn{2}{|c}{Anti-Sisi} \\\hline
% Hashtag & Count & Hashtag & Count \\ \hline
% UAE	&	5,253	&	Syria	&	10,400	\\
% KSA	&	3,151	&	KSA	&	8,813	\\
% Egypt	&	3,103	&	Egypt	&	7,085	\\
% Yemen	&	2,355	&	People\_Want\_To\_Overthrow\_The\_Regime	&	5,509	\\
% United\_Against\_Enemies\_Of\_Egypt	&	1,941	&	Turkey	&	5,024	\\
% Mohamed bin Zayed	&	1,892	&	Revolution\_Coming	&	4,351	\\
% Zayed\_Charity\_Marathon	&	1,637	&	Iran	&	4,344	\\
% Iran	&	1,619	&	Rassd (media)	&	3,615	\\
% 44th\_National\_Day (UAE)	&	1,501	&	Isis	&	3,218	\\
% UAE	&	1,335	&	Breaking\_News	&	3,160	\\
% % Iranian\_Political\_Hypocrisy	&	1,300	&	Qatar	&	3,012	\\
% % High\_Crown	&	1,273	&	Iraq	&	3,005	\\
% % Long\_Live\_Egypt	&	1,266	&	Sisi	&	2,963	\\
% % Know\_The\_Instegator	&	1,194	&	Our\_Revolution\_We\_Will\_Complete	&	2,488	\\
% % ISIS	&	1,193	&	Russia	&	2,408	\\
% % Mohamed\_bin\_Rashed	&	972	&	Depart\_Loser	&	2,393	\\
% % Dubai	&	886	&	Unity\_Group	&	2,387	\\
% % Emirates\_Jordanian\_Camp	&	810	&	Yemen	&	2,376	\\
% % Aden	&	765	&	We\_Will\_Expose\_You	&	2,022	\\
% % Sisi	&	765	&	Aleppo	&	1,881	\\
% \end{tabular}
% \caption{Top hashtags (translated) for pro- and anti-Sisi seminar users}
% \end{center}

% \end{table}

\begin{table}[ht]
\begin{center}
\scriptsize
\begin{tabular}{r|l|r||r|l|r}
\multicolumn{3}{c|}{Pro-Sisi} & \multicolumn{3}{|c}{Anti-Sisi} \\\hline
User	&	Description	&	Count	&	User	&	Description	&	Count	\\ \hline
Forsan\_UAE	&	UAE	user &	33,229	&	ALAMAWI	&	Syrian pro-Arab Spring media	&	58,203	\\
HenryKesnger	&	Unknown	&	10,452	&	almokhtsar	&	News aggregator 	&	29,968 \\
fdeet\_alnssr	&	KSA	&	7,817	&	raawnq	&	suspended	&	26,117	\\
meshaluk	&	KSA user in London	&	5,840	&	EHSANFAQEH	&	Jordanian author	&	23,005	\\
alkhaldi\_ksa	&	suspended	&	4,578	&	Wesal\_TV	&	KSA (anti-Iran media)	&	19,767	\\
Dhahi\_Khalfan	&	UAE user	&	4,258	&	S7K00	&	suspended	&	19,403	\\
bestwalid92	&	Egypt	&	3,778	&	AJArabic	&	Aljazeera (media)	&	18,212	\\
amirelghareb	&	Egypt user	&	3,696	&	kasimf	&	Aljazeera presenter	&	15,492	\\
a\_e5552000	&	suspended	&	3,627	&	AboShla5Libraly	&	 KSA user	&	14,389	\\
shereen\_hussen	&	Egypt user	&	3,502	&	YZaatreh	&	Palestinian author	&	13,639	\\
\end{tabular}

\caption{Top retweeted accounts for pro- and anti-Sisi seminar users.}
\label{topRetweetedAccounts}

\end{center}
\end{table}
\begin{figure}[h]
\begin{center}
\includegraphics[width=\linewidth]{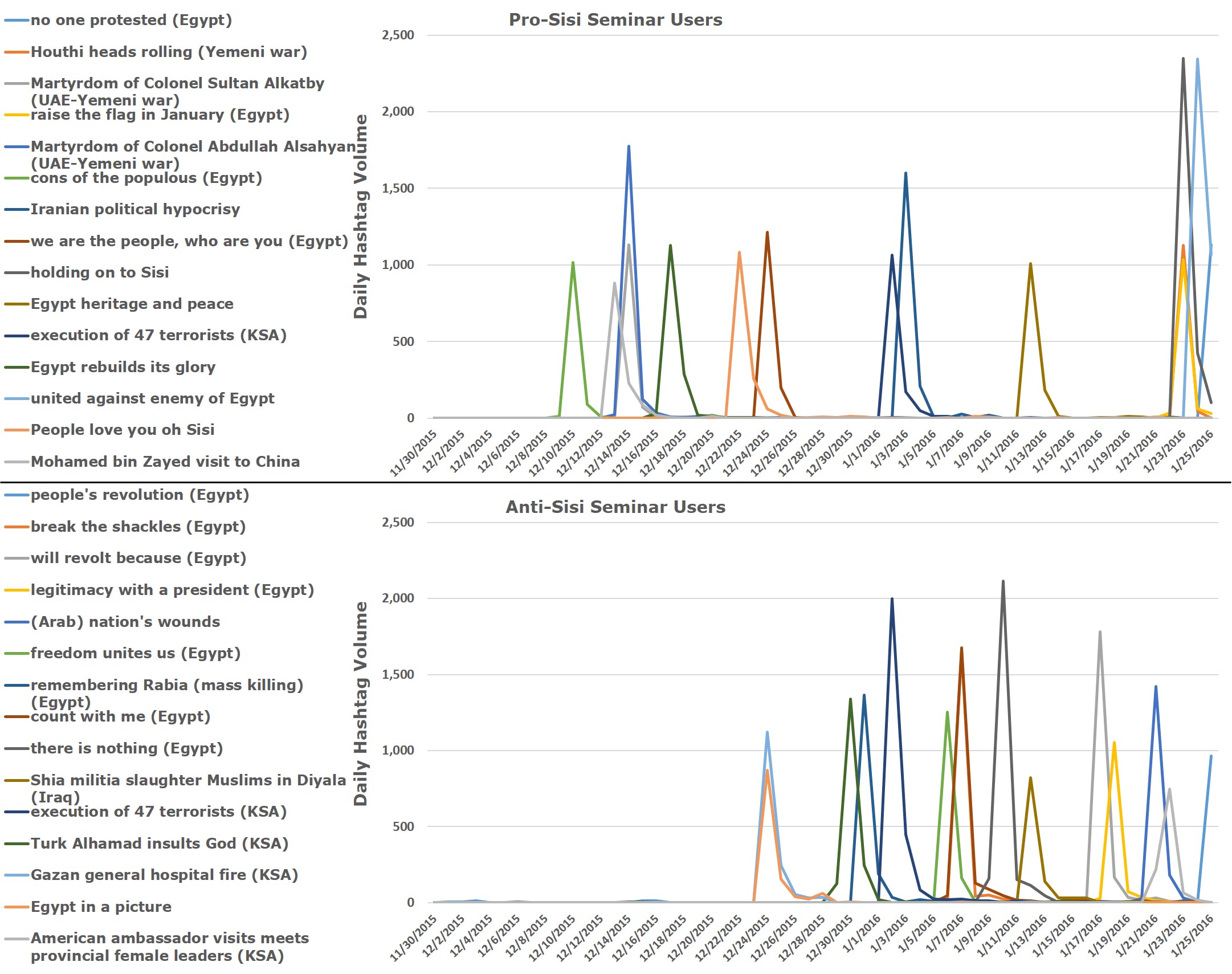}
\caption{Top hashtag campaigns for pro- and anti-Sisi seminar users.}
\label{top-hashtags-pos}
\end{center}
\end{figure}

\noindent Seminar users typically engage in campaigns to promote specific messages typically with associated hashtags.  We identified such campaigns by identifying hashtags that combine both high volume and high standard deviation ($\sigma$) from day to day. Thus, we scored hashtags by combining volume and standard deviation:
%using the following formula:

\begin{equation}
score = \frac{\sigma}{\sum daily\_counts}
\label{ranking-formula}
\end{equation}

\noindent We only considered hashtags that were used more than 100 and 300 times for pro- and anti-Sisi groups, respectively, and we ranked them using scoring formula (\ref{ranking-formula}). We picked different thresholds for the two groups in order to account for the difference in volume.
%between them. 
% Consequently, we considered hashtags from the pro- and anti-Sisi seminar users appearing more than 110 and 352 times, respectively. \textbf{Preslav: These are strange numbers to use as a threshold; what is the motivations to use these values. Why not 100? 1000?} 
Figure \ref{top-hashtags-pos} shows the top-15 hashtag campaigns for both groups.  We can see that the campaigns typically lasted for one or two days and then died out.  The pro-Sisi group was involved in continuous campaigns, while the anti-Sisi group campaigns started at the end of December as the anniversary of the 2011 Egyptian uprising drew closer. For the pro-Sisi group, 3 out of 15 of the top hashtags were related to Egypt and 7 were related to UAE.  In contrast, 14 out of 15 hashtags were related to Egypt in the anti-Sisi group.  

\subsection{Effectiveness, Cohesiveness, and Interactions between Seminar Users}

We set out to answer three research questions aimed to assess the effectiveness and the cohesiveness of seminar users, and the extent to which they interact with each other.  Our first question is the following: how successful are seminar users in penetrating the mainstream?  To answer this question, we considered the 5,618 pro- and anti-Sisi users that our classifier deemed as non-seminar users.  

We considered the top-100 most used hashtags % that accounted for at least 0.2\% of the total hashtag volume (or more than 4,558 times), 
with a score greater than 0.02, using formula~(\ref{ranking-formula}).  Then, we compared the top-100 hashtags for the pro- and anti-Sisi seminar groups to the top hashtags for non-seminar users, focusing on the following statistics: (\emph{i})~percentage of hashtags that made the list from non-seminar users, (\emph{ii})~average rank of hashtags in the list, and (\emph{iii})~volume magnification factor, which is the volume of hashtags from non-seminar users divided by the volume generated by seminar users.  Table~\ref{seminar-user-penetration} reports on these three statistics for the pro- and anti-Sisi seminar users.  We can see that anti-Sisi seminar users were more successful than pro-Sisi ones in penetrating the mainstream with 29\% (compared to 10\%) of their top-100 hashtags appearing in the top hashtags for non-seminar users with an average rank of 22.6 (vs. 46.1). However, volume magnification for pro-Sisi users is much higher (9.0 vs. 1.6).  

\noindent Three caveats need to be taken into account when reading these results: (\emph{i})~anti-Sisi seminar users outnumber pro-Sisi ones by a factor 2.7 to 1, (\emph{ii})~pro-Sisi seminar users do not concentrate on Egypt-related issues, and (\emph{iii})~seminar users may adopt an existing hashtag, instead of introducing a new one.  

\begin{figure}[ht!]
\begin{center}
\includegraphics[width=.95\linewidth]{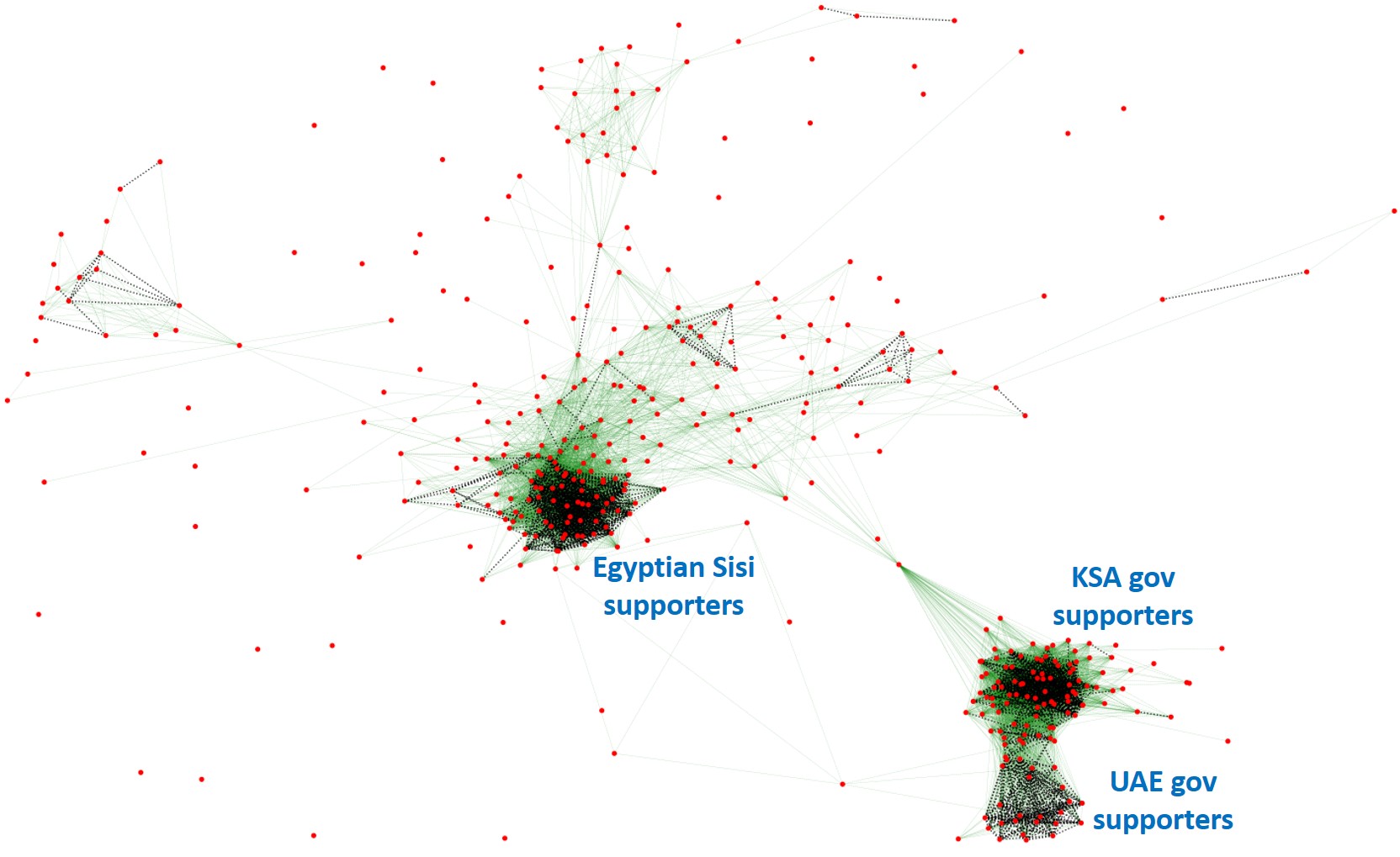}\\
\includegraphics[width=.95\linewidth]{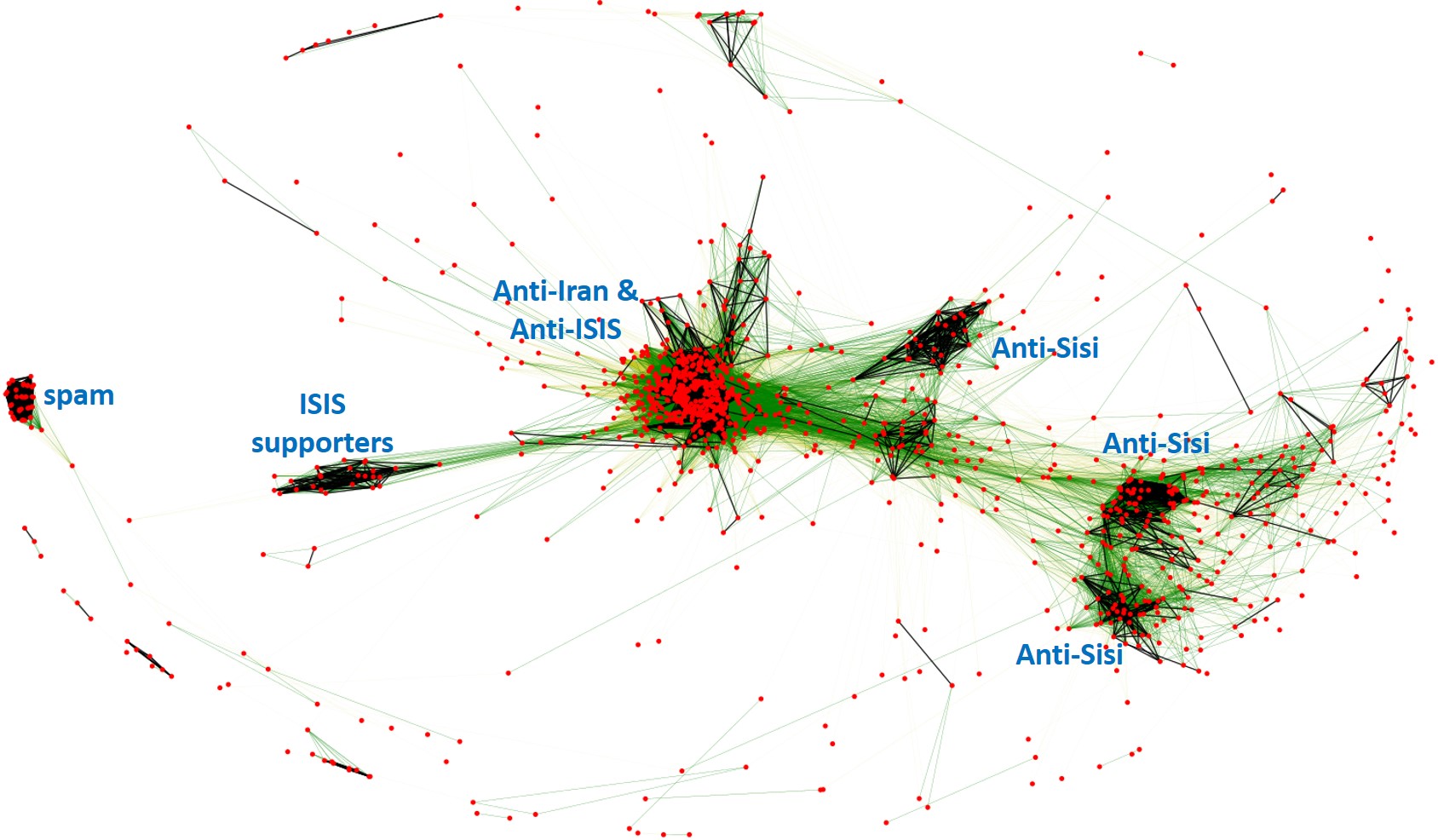}
\label{user-network-graph}
\caption{Pro- and anti-Sisi user network graph (top and bottom, respectively). We indicate cosine similarity greater than 0.8 as black edges, between 0.6 and 0.8 as green edges, and less than 0.6 as light yellow edges. We further show text labels for the main strongly-connected cliques.}
\end{center}
\end{figure}

\noindent Our second research question we study is the following: how similar are seminar users in each camp?  To answer this question, we computed the cosine similarity between pro- and anti-Sisi seminar users using the frequency of the hashtags that they used as features.  Next, we drew the network of users shown in Figure~\ref{user-network-graph} using NetworkX,\footnote{\url{http://networkx.readthedocs.io/en/networkx-1.10/}} which uses Fruchterman-Reingold force-directed algorithm to space out the nodes.  As the network graphs show, the seminar users are not homogeneous, but rather they are groups of strongly-connected cliques. We labeled the major cliques in the graphs based on sample users from each clique. We can see that pro-Sisi seminar users are composed of three major cliques from Egypt, KSA, and UAE. %, of which there 4 different pro-Sisi groups and 1 pro-UAE government group.  
Given the apparent skew toward UAE-related topics, it seems that the pro-UAE group is most productive in terms of volume.  For the anti-Sisi seminar users, there is a large anti-Iran group, which is also pro-Syrian revolutionaries, there are three different Egyptian anti-Sisi groups, one pro-Islamic State (ISIS) group, and one group of spammers.  All of these groups are supportive of the Arab Spring uprisings in one way or another. Prior work has discussed the relationship between support for ISIS and the Arab Spring \cite{magdy2016failedrevolutions}.

Our third research question is the following: how much do the two groups of seminar users interact? To answer this question, we measured %two things, namely: are users following others in the opposing camp; and 
whether the two camps used the same hashtags.  %Table \ref{cross-follow} shows the number of times a seminar user follows another user in his camp or the opposing camp.  The results in the table suggest that seminar users often follow others in their camp, and rarely follow others in the opposing camp. For hashtags, we wanted to measure how often one side would use hashtags of the opposing sides.  Table \ref{cross-hashtag} shows the percentage of top hashtags for either side that are being used by the other side at different penetration levels, which is the percentage of times the other side used the hashtag. 
In essence, we wanted to detect hashtag hijacking.  We found that the two camps shared only nine of the top 100 most frequent hashtags, and the rest were used exclusive by one of them. This included seven shared hashtags of broad interest such as \textit{\#KSACutsDiplomaticTiesWithIran} and \textit{\#DubaiFire}.  The other two hashtags were \textit{\#PeopleWantToOverthrowRegime} (used 96\% of the time by the anti-Sisi camp) and \textit{\#KnowThePerpetrator} (used 65\% of the time by the pro-Sisi camp).
% As the table shows, the vast majority of hashtags were exclusively used by one side or the other.  We also manually inspected the top hashtags with high penetration levels (more than 20\%), and most of the hashtags were generic hashtags, such country names and group names.  The only exceptions were that hashtags ``no\_one\_protested''\footnote{ one protested in the January 25 anniversary} and ``manly\_Sameh\_Shoukry''\footnote{referring to Egyptian foreign minister}, which originated from pro-Sisi seminar users where 29\% and 33\% of the volume respectively originated from anti-Sisi seminar users.  
The results suggest that both pro- and anti-Sisi seminar users operate fairly independently and with limited interaction.

% \begin{table}
% \begin{center}
% \begin{tabular}{r|r|r}
% 			& \multicolumn{2}{c}{following} \\ \hline
% follower			& pro-Sisi 	& anti-Sisi 	\\ \hline
% pro-Sisi 	& 988				& 6		\\
% anti-Sisi	& 10				& 5,130		\\
% \end{tabular}
% \label{cross-follow}
% \caption{Cross follow relationship between pro- and anti-Sisi seminar users}
% \end{center}
% \end{table}

% \begin{table}
% \begin{center}
% \begin{tabular}{r|r|r}
%  			& \multicolumn{2}{r}{Percentage of Top hashtags} \\ \hline
% Penetration Level	& Pro-Sisi & Anti-Sisi \\ \hline
% $>$ 0.4	& 8\%	& 2\% \\
% $>$ 0.3	& 1\%	& 2\% \\
% $>$ 0.2	& 0\%	& 2\% \\
% $>$ 0.1	& 0\%	& 2\% \\
% $>$ 0.0	& 0\%	& 1\% \\
% $=$ 0.0	& 91\%	& 91\% \\
% \end{tabular}
% \label{cross-hashtag}
% \caption{Cross hashtag usage between pro- and anti-Sisi seminar users}
% \end{center}
% \end{table}

\begin{table}
\begin{center}
\begin{tabular}{r|c|c|c}
			& \% of appearance 	& Avg. Rank 	& Vol. magnification \\ \hline
pro-Sisi 	& 10\%				& 46.1		& 9.0x				\\
anti-Sisi	& 29\%				& 22.6		& 1.6x				\\
\end{tabular}
\label{seminar-user-penetration}
\caption{Penetration of pro- and anti-Sisi seminar users into the mainstream.}
\end{center}
\end{table}

% \subsection{Organization}

% { \color{blue} And these users are colluding/organizing (do clustering to find users behaving suspiciously close)}

% \subsection{Effectiveness}

% { \color{blue} And these users are actually effective in spreading their ideas (back up with data, how many of the top hashtags they are responsible for)}
% \vspace{-10pt}

\section{Discussion and Conclusions}
\label{sec:conclusion}

We have introduced the so-called % a type of Twitter users that we refer to as 
``seminar users'', as users who engage in political propaganda. The existence of such users can distort social media analysis and can be socially and politically troubling. % , as it is sometimes difficult to discern the motives of political actors behind these accounts. Latest research on the use of Twitter in politics suggests the volume of such communication may be predicted by money spent in a campaign \cite{mcgregor2017twitter}.  
Thus, we have presented a robust method for automatically detecting such users by looking at their interactions, their tweet content diversity, and their tendency to use sentiment-bearing and offensive words.  Note that we did not use user similarity features as mining the user graph of interactions is computationally expensive. 

\noindent Our seminar vs. regular user classifier achieved precision of 84.4\% and recall of 76.1\% on our dataset, which is a very strong result. 
We have further demonstrated that seminar users are quite different from bots. Complementary to recent research that has shown that it is possible to distinguish automatically real human Twitter accounts from bots \cite{stieglitz2017social}, our work points to a need for detecting networks of colluding Twitter users, beyond the detection of ephemeral bot accounts.
 %\textcolor{red}{More on how well our classifier does against, say, bot classification?}

We further applied our classifier on Twitter users who discuss a politically controversial topic, namely attitude toward General Sisi of Egypt. Our classifier labeled 25\% of these users as seminar users, but we suspect that the actual percentage should be even higher. Moreover, our analysis shows that seminar users produce more tweets on average compared to normal users, e.g., more than half of the pro-Sisi and about a third of the anti-Sisi tweets in our dataset were posted by seminar users. We further found that seminar users are often successful in affecting the mainstream discourse, which is evident by the hashtags that they popularize. The prevalence of seminar users in the Arabic Twitter-sphere and the large volume of tweets that they produce complicate Twitter-based studies, particularly for controversial or political issues. The success of seminar users in affecting the mainstream has the potential of promoting human rights abuses, or making such abuses sound like a normal thing. Thus, the ability to detect such users can help analysis in two major ways, namely (\emph{i})~filtering out such users can help focus analysis on normal users, and (\emph{ii})~focusing exclusively on seminar users can help elucidate what political actors are promoting aside from what they say officially. 

We have shown that seminar users with similar stances on specific issues may actually belong to wildly different or even opposing groups of Twitter users, and that they may even come from different countries altogether.  In particular, for anti-Sisi seminar users, we can clearly see a group that supports ISIS and another larger group that opposes ISIS. In the example of pro-Sisi seminar users, we have seen such users from UAE and KSA, who work across borders to support Sisi and to oppose his adversaries.  Moreover, we have shown that groups of seminar users may or may not engage with the hashtags of opposing groups, e.g.,~in the form of hashtag hijacking. Overall, in our case study, we have found very limited evidence for this kind of cross-group engagement. We suspect that the seminar users phenomenon is not unique to the Arab region, but exists in different regions of the world with varying potency.  We plan to test whether our seminar vs. normal users classifier % of seminar users trained on data from the Arab region 
can generalize to other regions of the world, and whether seminar users %.  In doing so, we can ascertain if such groups 
behave similarly across regions and whether they share some universal characteristics.

Last but not least, by focusing our study on the political issues in the Arab world, we contribute to a growing literature examining the role of social media in political communication around the world. Although studies about United States are most prominent (e.g.,~those on the Presidential elections \cite{wells2016trump}), there is also recent work focusing on Latin America \cite{waisbord2017populist}, China \cite{song2016not}, Europe \cite{cross2016tweeting}, and the Middle East \cite{magdy2016failedrevolutions}. We hope that the present work would enable further research on the role and use of social media in Arab politics.  % and The Arabic language tool pipeline we present here, as well as local insights of this study open another locale of analysis.

% in our set authored 5.1 million tweets compared 9.9 million for non-seminar users, meaning that they accounted for more than a third of the tweet volume on the topic.  

% Discuss fine distinction from bot-based astroturfing.

% Discuss peculiarities of Arab NLP needed for this (as opposed to current English tools). Link to our tools.

% Discuss peculiarities of Arab (Egypt?) political scene (as compared to Western).

% Discuss repercussions for Twitter-based political analysis (and generic political discourse analysis).

\bibliographystyle{splncs03} 
\bibliography{bibliography,bib}
% \begin{thebibliography}{5}
% %
% \bibitem {clar:eke}
% Clarke, F., Ekeland, I.:
% Nonlinear oscillations and
% boundary-value problems for Hamiltonian systems.
% Arch. Rat. Mech. Anal. 78, 315--333 (1982)

% \bibitem {clar:eke:2}
% Clarke, F., Ekeland, I.:
% Solutions p\'{e}riodiques, du
% p\'{e}riode donn\'{e}e, des \'{e}quations hamiltoniennes.
% Note CRAS Paris 287, 1013--1015 (1978)

% \bibitem {mich:tar}
% Michalek, R., Tarantello, G.:
% Subharmonic solutions with prescribed minimal
% period for nonautonomous Hamiltonian systems.
% J. Diff. Eq. 72, 28--55 (1988)

% \bibitem {tar}
% Tarantello, G.:
% Subharmonic solutions for Hamiltonian
% systems via a $\bbbz_{p}$ pseudoindex theory.
% Annali di Matematica Pura (to appear)

% \bibitem {rab}
% Rabinowitz, P.:
% On subharmonic solutions of a Hamiltonian system.
% Comm. Pure Appl. Math. 33, 609--633 (1980)

% \end{thebibliography}

\end{document}